\documentstyle[aps,preprint]{revtex}

\newcommand{\be}{\begin{equation}}
\newcommand{\ee}{\end{equation}}
\newcommand{\bea}{\begin{eqnarray}}
\newcommand{\eea}{\end{eqnarray}}
\newcommand{\nn}{\nonumber \\}
\newcommand{\HH}{{\cal H}}
\newcommand{\HHi}{{\cal H}_i}
\newcommand{\B}{{^3\!{B_{\scriptscriptstyle +}}}}
\newcommand{\BB}{{^3\!{B_{\scriptscriptstyle -}}}}
\newcommand{\0}{{^3\!{B_{\scriptscriptstyle 0}}}}
\newcommand{\1}{{^3\!{B_{\scriptscriptstyle 1}}}}
\newcommand{\two}{{^2\!{B_{\scriptscriptstyle +}}}}
\newcommand{\bb}{{^2\!{B_{\scriptscriptstyle -}}}}
\newcommand{\2}{{^2\!{B_{\scriptscriptstyle 0}}}}
\newcommand{\3}{{^2\!{B_{\scriptscriptstyle 1}}}}

\newcommand{\munu}{{\mu\nu}}
\newcommand{\sss}{{\scriptscriptstyle}}
\begin{document}
\draft
\preprint{$\hbox to 5 truecm{\hfil Alberta-Thy-19-94}
\atop
\hbox to 5 truecm{\hfil gr-qc/9405041}$}
\title{\Large Eternal Black Holes and Quasilocal Energy \footnote{Based on the
talk presented by E.A. Martinez at the {\it  Lake Louise Winter School on
Particle Physics and Cosmology}, February 20-26, 1994.}}
\author{Valeri Frolov and Erik A. Martinez\footnote{electronic addresses:
frolov@phys.ualberta.ca, martinez@phys.ualberta.ca}}
\address{Theoretical Physics Institute,\\
Department of Physics, University of Alberta,\\
Edmonton, Alberta T6G 2J1, CANADA.}
\maketitle
\begin{abstract}
We present the gravitational action and Hamiltonian for a spatially bounded
region of an eternal black hole. The Hamiltonian is of the general form $H
=H_{+} - H_{-}$, where $H_{+}$ and $H_{-}$ are respectively the Hamiltonian for
the regions $M_+$ and $M_-$ located  in the left and right wedges of the
spacetime. We
construct explicitly the  quasilocal energy for the system and discuss its
dependence on the time direction induced at the boundaries of the manifold.
This paper extends the analysis of Ref.~[1] to spacetimes possesing a
bifurcation surface and two timelike boundaries. The  construction suggests
that
an interpretation of black hole thermodynamics based on thermofield dynamics
ideas can be generalized beyond perturbations to the gravitational field itself
of a bounded spacetime region.
\end{abstract}
\vfill\eject

Recent work on black hole thermodynamics  has suggested an
definition of quasilocal energy for spatially bounded systems in general
relativity~[1]. In thermodynamic applications  this energy plays the role of
the thermodynamical internal energy conjugate to inverse temperature~[2].
This quasilocal energy is obtained directly from the gravitational action and
Hamiltonian and includes negative contributions due to gravitational binding.

In the present paper we will present the action and Hamiltonian for an eternal
black hole and discuss in detail the  quasilocal energy for the system.
The motivation for studying the action of a fully extended black hole is at
least twofold:  in the first place, it is important to understand the general
properties of the boundary  Hamiltonians when the spacetime is bounded not by
one
but by two timelike boundaries separated by a bifurcation surface.
In the second place, the  action for eternal black holes will play a
fundamental
part in the study  and generalization to the full gravitational field of
interpretations of black hole thermodynamics based on the existence of
 spacetime regions $M_+$ and $M_-$.
The density matrix describing internal states of a black hole can be obtained
by tracing over degrees of freedom in $M_+$. The density matrix can be used to
calculate the entropy connected with the region $M_-$ (black hole entropy)~[3].
In this way pure states for
an eternal black hole would result in mixed thermal states for observers
located
on either disconnected region. One can relate  the density of states with
the action for an eternal black hole and consider its thermodynamical
implications. In this approach an interesting relation with thermofield
dynamics arises (for more details see Ref.~[4]).

We start by discussing the kinematics of a completely extended stationary
spacetime ($t^\mu$ denoting its Killing vector) with the topology of an eternal
black hole.
The corresponding Kruskal spacetime is the union of four sectors $R_+$, $R_-$,
$T_+$, and $T_-$~[5]. The regions $R_+$ and $R_-$ are asymptotically flat and
$t^\mu$ is timelike at their spatial infinities.
We concentrate attention in two wedges   $M_+$ and $M_-$ located in the right
($R_+$) and left ($R_-$) sectors of the Kruskal diagram.
Denote by $\Sigma_{\pm}$ the spacelike boundaries of $M_{\pm}$ and
by ${^3\!{B_{\scriptscriptstyle {\pm}}}}$ their timelike boundaries. The
regions
$M_+$ and $M_-$ intersect at a two-dimensional spacelike surface (bifurcation
surface).

The spacetime line element can be written in the general form
\be
 ds^2 = -N^2 dt^2 + h_{ij} (dx^i + V^i dt)(dx^j + V^j dt)\ ,
\ee
where $N$ is the lapse function. The four-velocity is defined
by $u_\mu = -N \, \partial_\mu t$,
where the lapse function $N$ is defined so that  $u \cdot u = -1$.
The four-velocity $u^\mu$ is the timelike unit vector that is  normal to the
spacelike hypersurfaces ($t = {\rm constant}$) denoted by $\Sigma_{t}$.
The symbol $\Sigma_{t\,(\pm)}$ indicates the part of $\Sigma_{t}$ located in
$M_{\pm}$. The Killing vector $t^\mu$ and the
four-velocity $u^\mu$  are related by
\be
t^\mu = Nu^\mu + V^\mu \ ,
\ee
so that
$V^{i}= h^i _0 = - N u^i$ is the shift vector.
We choose the spacelike boundaries to coincide with surfaces of fixed values of
$t$ so that the Killing time $t$ is the scalar function that uniquely
(everywhere outside the bifurcation surface) labels the foliation. The lapse
function $N$ is positive (negative) at $M_+$ ($M_-$) and
zero at the bifurcation point. With respect to the global time, $u^\mu$ is
future
oriented in $M_+$ and  past  oriented in  $M_-$. The spacelike normal $n^{i}$
to the three-dimensional boundaries ${^3\!{B}}$ is defined to be  outward
pointing at the boundary $\B$  and normalized so that   $n \cdot n=+1$. The
normal to the boundary $\BB$ is obtained by a continuous transport along
$\Sigma_{t}$ of the outward pointing normal to $\B$. The foliation  is
restricted
by the conditions $(u \cdot n)|_{{^3\!{B}}} = 0$.

In parallel to  Ref.~[1],
the metric and extrinsic curvature of $\Sigma$ as a surface embedded in
$M$ are denoted $h_{ij}$ and  $K_{ij} = -h_i ^{k} \nabla_{k}u_{j}$
respectively, while the metric and extrinsic curvature of the boundaries
${^3\!{B}}$ as  surfaces embedded in $M$ are $\gamma_{ij}$ and
$\Theta_{ij} = -\gamma_i ^{k} \nabla_{k} n_{j}$.
The intersection of the boundaries $\B$ and $\BB$ with  $\Sigma$ are
(topologically) spherical two-dimensional surfaces denoted $\two$ and $\bb$
respectively.
Finally, the induced metric and
extrinsic curvature of the boundaries ${^2\!{B}}$ as surfaces embedded on
$\Sigma$ are  ${\sigma}_{ab}$ and
$k_{ab} = -\sigma_{a}^{k} D_{k} n_{b}$ respectively. The simbols $\nabla$ and
$D$ denote covariant derivatives with respect to $g_{\munu}$ and
$h_{ij}$.  The traces of the extrinsic curvatures $\Theta_{ij}$ and
$k_{ij}$ are related by $\Theta = k - n_i a^i $,
where  $ a^\mu =  u^\alpha \nabla _{\alpha} u^\mu $ is the acceleration of the
timelike normal $u^{\mu}$.

The covariant form of the gravitational action for the spacetime described
above is discussed in detail in Ref.~[4] for different choices of boundary
conditions. The appropriate action when the three-geometry is fixed at the
boundaries of $M$ is
\bea
 S &=&
{1\over2\kappa} \int_{M_+} d^4 x \,\sqrt{-g}\,\Re  + {1\over\kappa}
\int_{({\scriptscriptstyle +})t'}^{t''} d^3 x \,\sqrt{h}\, K
 - {1\over\kappa} \int_{\B} d^3x \,\sqrt{-\gamma}\, \Theta \nn
&-&{1\over2\kappa} \int_{M_-} d^4 x \,\sqrt{-g}\,\Re
+ {1\over\kappa} \int_{({\scriptscriptstyle -})t'} ^{t''} d^3 x \,\sqrt{h}\, K
- {1\over\kappa} \int_{\BB} d^3 x \,
\sqrt{-\gamma}\, \Theta \label{bigaction}
\eea
where  $\Re$ denotes the four-dimensional scalar curvature and the notation
$\int_{({\scriptscriptstyle +})t'}^{t''}$  represents an  integral over the
three-boundary $\Sigma_{+}$ at $t''$ minus an integral over the three-boundary
$\Sigma_{+}$ at $t'$. Units are chosen so that $G =c = \hbar =1$, and  $\kappa
\equiv 8 \pi$. The relative signs of the terms in (\ref{bigaction}) are
obtained
by a careful application of Stokes theorem to a spacetime with a bifurcation
surface~[4]. The action (\ref{bigaction}) has to be complemented with a
subtraction term~[6,1]. This term is discussed  in Ref.~[4] but
its effects on the results are discussed below.

We  write the Hamiltonian form of the action following the standard
procedure by recognizing that we have a global direction of time at the
boundaries inherited by the time vector field $t^\mu $ defined
at ${^3\!{B}}$.  The momentum $P^{ij}$ conjugate
to the three-metric $h_{ij}$ of $\Sigma$ is defined  as
\be
P^{ij} = {1\over2\kappa} \sqrt{h} (K h^{ij} - K^{ij}) \ ,
\ee
and the definition of $K_{\mu \nu}$ implies that
\be
K_{ij} = -\nabla _{i} u_j =
-{1\over {2N}} \bigl[ \dot h_{ij} - 2D_{(i} V_{j)}
\bigr]\ , \label{Kij}
\ee
where the dot denotes differentiation with respect to the  time $t$.
In terms of the quantities~[1,2]
\be
\varepsilon \equiv  \bigl( k/\kappa\bigr) \label{edensity} \ , \,\,
 j^i \equiv -2  \sigma^i_{k} n_{\ell}P^{k\ell}/\sqrt{h}\label{mdensity}
\ee
the action can be written in the desired Hamiltonian form
\be
 S = \int_M d^4x  P^{ij} \dot h_{ij} - \int H dt \ ,\label{Hamaction}
\ee
where  the gravitational Hamiltonian $H$ is given by
\be
H =\int_{\Sigma} d^3x \bigl(N\HH + V^i \HHi \bigr)
+ \int_{\two} d^2x \sqrt{\sigma} \bigl(  N \varepsilon - V^i  j_i\bigr)
 - \int_{\bb} d^2x \sqrt{\sigma} \bigl( N \varepsilon - V^i  j_i\bigr)\ .
\label{Ham}
\ee
The Hamiltonian has both volume and  boundary contributions. The volume
part involves the Hamiltonian and momentum constraints
\bea
\HH &=& (2\kappa) G_{ijk\ell}\, P^{ij}\, P^{k\ell} - \sqrt{h}
      \, R/(2\kappa) \ , \nn
\HHi &=& -2 D_j \, P_i^j \ .
\eea
where
$G_{ijk\ell} = (h_{ik} h_{j\ell} + h_{i\ell} h_{jk} - h_{ij} h_{k\ell})
/(2\sqrt{h}) $ and $R$ denotes the scalar curvature of the surfaces $\Sigma$.
Notice that the action (\ref{Hamaction}) has its only  non-zero contribution
from the boundary terms of the Hamiltonian~[2]. The first term of
(\ref{Hamaction}) is zero by stationarity  and the volume part of the
Hamiltonian (\ref{Ham}) is  equally zero when the Hamiltonian and
momentum constraints are imposed. The  Hamiltonian  can be written in the
suggestive form
\be
 H =H_{+} - H_{-} \ , \label{sub}
\ee
where the Hamiltonian $H_+$ of $M_+$ and the Hamiltonian $H_-$ of $M_-$ are
\bea
H_{+} &=&\int_{\two} d^2x \sqrt{\sigma} \bigl( N \varepsilon - V^i  j_i \bigr)
\nn
H_{-} &=& \int_{\bb} d^2x \sqrt{\sigma} \bigl( N \varepsilon - V^i  j_i \bigr)
 \ .\label{H+H-}
\eea
Following Ref.~[1], the quasilocal energy will be defined as the value of the
Hamiltonian that generates unit time translations orthogonal to the
two-dimensional boundaries. The  total quasilocal energy is consequently the
value of the Hamiltonian (\ref{sub}) (with $V_i = 0$) such that $|N| = 1$ at
both $\two$ and $\bb$, namely
 \bea
E_{\rm\sss tot} &=& E_{+} - E_{-}\ , \nonumber \\
E_{+}&=&\int_{\two} d^2x \sqrt{\sigma} (\varepsilon - {\varepsilon}^0 )
\nonumber \\
E_{-}&=& - \int_\bb d^2x \sqrt{\sigma} (\varepsilon - {\varepsilon}^0)\ .
\label{energy}
\eea
This is the quasilocal energy of a spacelike hypersurface
$\Sigma = \Sigma _+ \cup \Sigma _-$ bounded by two boundaries $\B$ and $\BB$
located in the two disconnected regions ${M_+}$ and ${M_-}$ respectively.  The
quantity $\varepsilon$ defined in terms of the trace $k$ of  extrinsic
curvature in (\ref{edensity}) is seen to be proportional to the surface energy
density of the gravitational field. We have included in these expressions the
relevant subtraction terms $\varepsilon ^0$ for the energy. The quantity
${\varepsilon}^0$ is defined by (\ref{edensity})  with $k$ replaced by
the trace $k^0$ of  extrinsic curvature corresponding to embedding  the
two-dimensional boundaries $\two$ and $\bb$ in three-dimensional Euclidean
space.
The delicacies involved in embedding a spacetime with two boundaries are
discussed in Ref.~[4].

To gain some intuition about the values of $E_+$ and $E_-$  consider
as an example a static Einstein-Rosen bridge with metric:
\be
ds^2 = -N^2 dt^2 + h_{yy} dy^2 + r^2 (y) d\Omega ^2 \ ,
\ee
where $N$, $h_{yy}$, and $r$ are functions of the  radial coordinate
$y$ continuously defined on $M$. The boundaries $\two$ and $\bb$ are located at
coordinate values $y=y_+$ and $y=y_-$ respectively. The normal to the
boundaries
is $n^ \mu = { (h^{yy})}^{1/2} \delta ^\mu _y $. Since this normal is defined
continuously along $\Sigma$, the value of $k$ depends on the function $r_{,y}$,
which is positive for $\two$ and negative for $\bb$. This results in
a quasilocal energy
\bea
E &=& E_+ - E_-  \nonumber \\
  &=& \bigg( r \big| r_{,y}\big| \Big[ 1 - (h^{yy})^{1/2} \Big] \bigg)_{y =
y_+}
- \bigg( r \big|r_{,y}\big| \Big[ 1 - (h^{yy})^{1/2} \Big] \bigg)_{y = y_-} \ .
\eea
It is easy to see that when the constraints are solved  $E_+ $ and $E_-$ tend
individually to the ADM mass $\cal M$ when the  boundaries $\B$ and $\BB$ tend
respectively to  right and left spatial infinities.
Notice that  the energies  $E_+$ and $E_-$ have the
same overall sign. In particular, the total enery is  zero for  boundary
conditions symmetric with respect to the bifurcation surface.
The dynamical aspects of the gravitational theory for spherically symmetric
eternal black holes was recently considered by Kucha\v{r}~[7]. Our expression
(\ref{H+H-}) reduces to his results when the black hole is spherically
symmetric and both boundaries $\B$ and $\BB$ are taken to their corresponding
spatial infinities.

The previous example illustrates a general property of the quasilocal energy
(\ref{energy}): the relative sign of  $E_+$ and $E_-$ is the same,
while the total energy is given by the subtraction $E_+ - E_-$. The same
property holds  for  the Hamiltonian. This is physically very
attractive: the regions $M_+$ and $M_-$ of an eternal black hole are
symmetric to each other, the time along the respective boundaries being
determined by the Killing time. The   energies associated to
$\B$ and $\BB $ are symmetric and the total energy reflects the
orientation reversal of the two boundaries.

The time direction induced at the boundaries $\B$ and $\BB$  considered above
was inherited from the Killing vector $t^\mu$. The value of the total action is
 independent of  time orientations at the boundaries. However, the
Hamiltonian does depend on this orientation.
If we prefer to use at the boundary $\BB$ the future directed $t_{-} = -t$, the
total boundary action takes the form
\be
S_B = -\int (H_+ dt_+ + H_- dt_-)  \ ,
\ee
where $t_+ \equiv t$ and
$H_+$ and $H_-$ are given by (\ref{H+H-}). The corresponding Hamiltonian $H =
H_+ + H_-$ determines the evolution of the complete system in future physical
time.

In this paper we are only concerned with the definition of quasilocal energy.
However, as shown  in Ref.~[1] for a spacetime with only one boundary, one can
obtain charge-like quantities from the gravitational action (\ref{Hamaction}).
For example,  the total angular momentum is given by $J_{\rm\sss tot} = J_{+} -
J_{-}$, where $J_+$ and $ J_{-}$ are proper integrals (over $\two$ and $\bb$
respectively) involving the momentum density $j^i$  of
the spacelike hypersurface defined by (\ref{mdensity}). We direct the reader to
Ref.~[4] for a complete discussion on the definition of angular momentum for an
eternal black hole.

It is interesting to compare the quasilocal energy (\ref{energy}) with the
quasilocal energy of a spacelike hypersurface $\Sigma$ bounded by two
boundaries $\0$ and $\1$ located both in ${M_+}$. In this case
\bea
E_{\rm\sss tot} &=& E_{0} - E_{1}\ , \nonumber  \\
E_{0}&=&\int_{\2} d^2x \sqrt{\sigma} (\varepsilon - {\varepsilon}^0 ) \ ,
\nonumber \\
E_{1}&=&  \int_{\3} d^2x \sqrt{\sigma} (\varepsilon -
{\varepsilon}^0)\ .  \label{e2inM+}
\eea
We emphasize that $E_1$ in (\ref{e2inM+}) and $E_-$ in (\ref{energy}) have the
same overall sign.  This is so because the extrinsic curvature $k$ has the same
sign  in both $\0$ and $\1$.

We  end with some remarks concerning the relation of the Hamiltonian $H =
H_+ - H_-$ for eternal black holes presented above with  Hamiltonians that
arises in applications of  thermofield dynamics.  Thermofield dynamics is a
theory of thermal fields based on  augmenting the physical Fock space $\cal F$
by a fictitious Fock space  ${\tilde {\cal F}}$ obtained from the first one by
a
conjugation operation~[8].  This results in a doubling of degrees of freedom
for
the system. A temperature dependent ``vacuum state" can be defined on the total
Fock space  ${\cal F} \otimes {\tilde {\cal F}}$ in such a way that vacuum
expectation values of any physical operator agree with their statistical
average
for an ensemble in thermal equilibrium. A pure state in  ${\cal F} \otimes
{\tilde {\cal F}}$  will then correspond to a mixed state in $\cal F$. The
Hamiltonian  for this extended  system has always the general form $ H = H_+ -
H_- $ where $H_+$ is the Hamiltonian associated to $\cal F$ and  $H_-$ the
Hamiltonian associated to $\tilde {\cal F}$.

In the original formulation of thermofield dynamics the space
$\tilde {\cal F}$ was merely formal. However, Werner Israel~[9] has suggested
that thermal effects of black holes can be explained by a
thermofield dynamics mechanism. For black
holes the space  ${\cal F} \otimes {\tilde {\cal F}}$  is
not necessarily formal but correspond to a fully extended (that is, eternal)
black hole. While the analysis in Ref.~[9]  was based  only on the study of
particle modes propagating on an eternal black hole background,  we believe
that
this interpretation  can be generalized beyond small perturbations
to the strong gravity regime, when all the quantities and in particular the
Hamiltonian refer to the  gravitational field itself.
However, this generalization has to reflect the fact that
black hole thermodynamics can only be rigourously defined for finite size
systems~[10,2].
It is then natural to expect that a full
generalization of the thermofield dynamic ideas to black holes should start
with the study of geometrodynamics of spatially bounded eternal black holes.
The
present paper is a step in this direction. We have shown that it is possible to
define the total gravitational Hamiltonian for a bounded region of an eternal
black hole.   This Hamiltonian is encoded in the boundary conditions and is
precisely of the thermofield dynamics form  $H = H_+ - H_- $. The Hamiltonian
was
calculated for a general spacetime, independently of spatial symmetries or
approximations. It only required that the spacetime had a horizon that
self-intersects in a bifurcation surface. The thermodynamics of eternal black
holes and the role of thermofield dynamics are  treated in detail in Ref.~[4],
where an explicit construction of the thermal properties of the system based on
the Hamiltonian $H = H_+ - H_- $ is presented.

\acknowledgements

This research was supported in part by the National Science and Engineering
Research Council of Canada. E. M. wishes to thank Werner Israel for his
continuous support, as well as the organizers of the Lake Louise Winter
Institute for their hospitality.
\vfill\eject

\vspace{7mm}
\centerline{\Large References}
\vspace{7mm}

\noindent
1. J. D. Brown and J. W. York, Jr., {\it Phys. Rev.} {\bf D47} (1993) 1407

\noindent
2. J. D. Brown, E. A. Martinez, and J. W. York, Jr., {\it Phys. Rev. Lett.}
{\bf 66}  (1991) 2281

\noindent
3. V. Frolov and I. Novikov, {\it Phys. Rev.} {\bf D48} (1993) 4545

\noindent
4. V. Frolov and E. A. Martinez, to be submitted to Phys. Rev. D, 1994.

\noindent
5. C.W. Misner, K. S. Thorne and J. A. Wheeler, {\it
Gravitation}, W. H. Freeman, 1973.

\noindent
6. G. W. Gibbons and S. W. Hawking, {\it Phys. Rev.} {\bf D15 }  (1977) 2752

\noindent
7. K. V. Kucha\v{r}, preprint, UU-REL-94/3/1, gr-qc/9403003

\noindent
8. H. Umezawa and Y. Takahashi, {\it Collective Phenomena} {\bf 2} (1975) 55

\noindent
9. W. Israel, {\it Phys. Lett.} {\bf 57A} (1976) 107

\noindent
10. J. W. York, Jr., {\it Phys. Rev.} {\bf D33} (1986) 2092

\end{document}